\documentclass[10pt, conference]{IEEEtran}
\usepackage{subfigure}
\usepackage{epsfig,graphicx,psfrag}
\usepackage{amsfonts,amsmath,amssymb}
\usepackage{bbm}

\newtheorem{theorem}{Theorem}[section]
\newtheorem{lemma}[theorem]{Lemma}

\begin{document}

\title{The Likelihood Encoder for Source Coding}

\author{
\authorblockN{Paul Cuff and Eva C. Song -- Princeton University}
}

\maketitle

\begin{abstract}
The likelihood encoder with a random codebook is demonstrated as an effective tool for source coding.  Coupled with a soft covering lemma (associated with channel resolvability), likelihood encoders yield simple achievability proofs for known results, such as rate-distortion theory.  They also produce a tractable analysis for secure rate-distortion theory and strong coordination.
\end{abstract}


\section{Introduction}

Here we present the simple analysis that results from using a likelihood encoder for source coding to replace, for example, a joint-typicality encoder (see \cite{cover-thomas}).  As with the joint-typicality encoder, the likelihood encoder is defined with a specific joint distribution in mind $P_{X,Y}$.  To encode a sequence $x_1,x_2,...,x_n$ (i.e. $x^n$) using a codebook of sequences $y^n(m)$, the encoder stochastically chooses an index $m$ according to the likelihood of $y^n(m)$ passed through the memoryless ``test channel'' $P_{X|Y}$.  That is,
\begin{eqnarray}
  \label{eq:likelihood encoder}
  P_{M|X^n}(m|x^n) & \propto & \prod_{t=1}^n P_{X|Y} (x_i|y_i(m)).
\end{eqnarray}

This encoder has been used by Cuff et. al. to achieve strong coordination in \cite{cuff-08, cuff-permuter-cover-10, cuff-ittrans13, satpathy-cuff-isit-13} and for secrecy systems in \cite{cuff-10-globecom}, \cite{cuff-10-allerton}, \cite{schieler-cuff-isit-13}, \cite{cuff-13-itw} because of the simplicity of the distribution that it induces.  In those scenarios, a precise understanding of the induced distributions is requisite for analysis.

Here we demonstrate that the likelihood encoder can be used in more traditional source coding problems, such as the achievability proof for rate-distortion theory.  An extension for the proof of source coding with side information at the receiver \cite{wyner-ziv} is straightforward.  Notice that the likelihood encoder is stochastic.  Unlike strong coordination or secrecy systems, most traditional rate-distortion settings can be shown not to benefit from a stochastic encoder.  In fact, a stochastic encoder is usually strictly suboptimal.  But just as the joint-typicality encoder is suboptimal, the motivation for using the likelihood encoder in these settings is in the analysis.

As an added curiosity, the likelihood {\em decoder} has recently been proposed for channel coding in \cite{yassaee-aref-gohari-13} because of its simple analysis.

The technique for analyzing the likelihood encoder relies on a soft covering lemma, analogous to the way that the joint-typicality encoder relies on the asymptotic equipartition principle.

\begin{lemma}[Soft Covering - {\cite[Lemma~IV.1]{cuff-ittrans13}}]
  Given a joint distribution $P_{X,Y}$, let ${\cal C}$ be a random collection of sequences $Y^n(m)$, with $m = 1,...,2^{nR}$, each drawn independently and i.i.d. according to $P_Y$.  Denote by $P_{X^n}$ the output distribution induced by selecting an index $m$ uniformly at random and applying $Y^n(m)$ to the memoryless channel specified by $P_{X|Y}$.  Then if $R > I(X;Y)$,
  \begin{eqnarray}
    {\mathbb E}_{\cal C} \left\| P_{X^n} - \prod_{t=1}^n P_{X} \right\|_{TV} & \to & 0
  \end{eqnarray}
  as $n$ goes to infinity, where $\| \cdot \|_{TV}$ is total variation.
\end{lemma}

The concept of the soft covering lemma was introduced by Wyner in \cite{wyner-75}, though with technical differences.  Also, this lemma plays a key role in the proof of the resolvability of a channel in \cite{han-verdu-93}.

Armed with the soft covering lemma, we now give a simple achievability proof for rate-distortion theory.

\section{Rate-Distortion Theory}

This theory addresses the optimal lossy compression of an i.i.d. source sequence $X^n$ distributed according to $X_i \sim \overline{P}_X$ under the following rate and fidelity constraints:

Encoder $f_n: \mathcal{X}^n \mapsto \mathcal{M}$ (possibly stochastic).

Decoder $g_n: \mathcal{M} \mapsto \mathcal{Y}^n$ (possibly stochastic).

Compression rate: $R$, i.e. $|\mathcal{M}|=2^{nR}$.

Fidelity requirement: $\mathbb{E} \; d(X^n, Y^n) \leq D$, where $d(x^n,y^n)=\frac1n \sum_{i=1}^nd(x_i,y_i)$.

Shannon's well-known theorem states that the infimum of achievable rates $R$ that can meet the distortion constraint $D$ ---optimized over $n$, $f_n$, and $g_n$---is given by
\begin{eqnarray*}
R &=& \min_{P_{Y|X}: \mathbb{E}[d(X,Y)]\leq D} I(X;Y). \label{rdfunction}
\end{eqnarray*}

We give the achievability proof. \\

\begin{proof}

Select $\overline{P}_{X,Y} = \overline{P}_{X} \overline{P}_{Y|X}$ such that $R > I(X;Y)$ and $\mathbb{E}[d(X^n, Y^n)] < D$.  Generate a random i.i.d. codebook according to $\overline{P}_Y$ and apply the likelihood encoder of \eqref{eq:likelihood encoder} with respect to $\overline{P}_{X,Y}$ as $f_n$.  Let $g_n$ simply produce the sequence $y^n(m)$ from the codebook.

Denote two distributions on the pair $(X^n,Y^n)$.  One is the distribution induced by the encoding and decoding, $P_{X^n,Y^n}$, which we wish to analyze.  The other is an idealized distribution $Q_{X^n,Y^n}$ given by
\begin{eqnarray*}
&& Q_{X^n,Y^n}(x^n,y^n) \\
&\triangleq& \left( \prod_{t=1}^n\overline{P}_{X|Y}(x_t|y_t) \right) \frac{1}{2^{nR}} \sum_{m=1}^{2^{nR}} \mathbbm{1}\{Y^n(m)=y^n\}.
\end{eqnarray*}

The distribution $Q_{X^n,Y^n}$ can be interpreted as a uniform distribution over the codebook applied to the memoryless channel given by $\overline{P}_{X|Y}$.  This is trivial to analyze.  Notice that when expectation is taken over the codebook,
\begin{eqnarray}
&&\mathbb{E}_{\mathcal{C}} \; Q_{X^n,Y^n}(x^n,y^n) \nonumber\\
&=& \left( \prod_{t=1}^n\overline{P}_{X|Y}(x_t|y_t) \right) \frac{1}{2^{nR}} \sum_{m=1}^{2^{nR}} \mathbf{P}_{\mathcal{C}}\{Y^n(m)=y^n\} \nonumber \\
&=& \prod_{t=1}^n \overline{P}_{X,Y}(x_t,y_t). \label{expectation}
\end{eqnarray}

The key step is to show that $P_{X^n,Y^n} \approx Q_{X^n,Y^n}$ in total variation, and then use properties of total variation.  First,
\begin{eqnarray*}
  {\mathbb E}_{\cal C} \left\| P_{X^n} - Q_{X^n} \right\|_{TV} & \to & 0,
\end{eqnarray*}
due to the soft covering lemma, noting that $P_{X^n} = \prod_{t=1}^n \overline{P}_X$.

Also, notice that $P_{Y^n|X^n} = Q_{Y^n|X^n}$ by construction of the likelihood encoder.  Therefore,
\begin{eqnarray}
  {\mathbb E}_{\cal C} \left\| P_{X^n,Y^n} - Q_{X^n,Y^n} \right\|_{TV} & \to & 0. \label{tv small}
\end{eqnarray}

A well-known property of total variation gives:
\begin{eqnarray*}
  \mathbf{E}_{P} \; d(X^n,Y^n) & \leq & \mathbf{E}_{Q}  \; d(X^n,Y^n) + 2 d_{max} \|P - Q\|_{TV}.
\end{eqnarray*}

Finally, we average over codebooks to complete the existence argument:
\begin{eqnarray*}
  \mathbf{E}_{\cal C} \; \mathbf{E}_{P} \; d(X^n,Y^n) & \leq & \mathbf{E}_{\overline{P}}  \; d(X,Y) + 2 d_{max} \mathbf{E}_{\cal C} \|P - Q\|_{TV} \\
  & < & D
\end{eqnarray*}
for $n$ large enough, due to \eqref{expectation} and \eqref{tv small}.
\end{proof}

\section{Acknowledgments}

This work is supported by the National Science Foundation (grant CCF-1116013) and the Air Force Office of Scientific Research (grant FA9550-12-1-0196).


\begin{thebibliography}{1}

\bibitem{cover-thomas}
T. Cover and J. Thomas, ``Elements of Information Theory,'' {\em Wiley}, second edition, 2006.

\bibitem{cuff-08}
P. Cuff, ``Communication Requirements for Generating Correlated Random Variables,'' {\em ISIT}, 2008.

\bibitem{cuff-permuter-cover-10}
P. Cuff, H. Permuter, T. Cover, ``Coordination Capacity,'' {\em IEEE Trans. on Info. Theory}, 56(9), 2010.

\bibitem{cuff-ittrans13}
P. Cuff, ``Distributed Channel Synthesis,'' {\em to appear in IEEE Trans. on Info. Theory}, arXiv:1208.4415.

\bibitem{satpathy-cuff-isit-13}
S. Satpathy, P. Cuff, ``Secure Cascade Channel Synthesis,'' {\em ISIT}, 2013.

\bibitem{cuff-10-globecom}
P. Cuff, ``A Framework for Partial Secrecy,'' {\em Globecom}, 2010.

\bibitem{cuff-10-allerton}
P. Cuff, ``Using a Secret Key to Foil an Eavesdropper,'' {\em Allerton}, 2010.

\bibitem{schieler-cuff-isit-13}
C. Schieler, P. Cuff, ``Rate-distortion Theory for Secrecy Systems,'' {\em ISIT}, 2013.

\bibitem{cuff-13-itw}
P. Cuff, ``Secrecy in Cascade Networks,'' {\em ITW}, 2013.

\bibitem{wyner-ziv}
A. Wyner and J Ziv, ``The Rate-distortion Function for Source Coding with Side Information at the Decoder,'' {\em IEEE Trans. on Info. Theory}, 22(1), 1976.

\bibitem{yassaee-aref-gohari-13}
M. Yassaee, M. Aref, and A. Gohari, ``A Technique for Deriving One-Shot Achievability Results in Network Information Theory,'' {\em ISIT}, 2013.

%
\bibitem{wyner-75}
A. Wyner, ``The Common Information of Two Dependent Random Variables,'' {\em IEEE Trans. on Info. Theory}, 21(2), 1997.

\bibitem{han-verdu-93}
T. Han and S. Verd\'{u}, ``Approximation Theory of Output Statistics,'' {\em IEEE Trans. on Info. Theory}, 39(3), 1993.

\end{thebibliography}
\end{document}